%% file: main.tex
\documentclass[letterpaper,twocolumn,10pt]{article}

\input{packages}

\newif\ifdraft
\drafttrue

\begin{document}

\input{macros}

\date{}

\title{Concurrency Fuzzing of the Linux Kernel with eBPF}

\author{
{\rm Jiacheng Xu\thanks{Both authors contributed equally}}\\
Zhejiang University\thanks{Research conducted primarily at the National University of Singapore}\\
\and
{\rm Dylan Wolff\footnotemark[1]\ \ \thanks{Corresponding Author: \texttt{wolffdy0@gmail.com}}}\\
National University of Singapore
\and
{\rm Xing Yi Han}\\
National University of Singapore
\and
{\rm Jialin Li}\\
National University of Singapore
\and
{\rm Abhik Roychoudhury}\\
National University of Singapore
} 


\maketitle

\begin{abstract}
Concurrency is indispensable for modern software systems to meet performance and scalability demands, yet concurrency bugs remain notoriously difficult to detect and reproduce.
Controlled Concurrency Testing (CCT) mitigates this challenge by systematically exploring thread interleavings through scheduling control.
However, existing CCT approaches for OS kernels largely rely on external enforcement mechanisms, such as custom hypervisors or invasive kernel patches, resulting in substantial overhead and limited maintainability and extensibility.
In this work, we present \tool, the first kernel-native concurrency fuzzing framework that rethinks scheduling as a first-class exploration mechanism.
\tool introduces a novel CCT scheduler with temporal isolation scheduling and embeds programmable
scheduling policies directly into the kernel dispatch path via eBPF, enabling fine-grained control over thread interleavings without customized hypervisors or extensive kernel core modification.
In addition, \tool provides a preemption-safe instrumentation mechanism for injecting scheduling points at critical kernel events and incorporates a two-phase fuzzing workflow to jointly explore both sequential and concurrent behaviors.
Our evaluation demonstrates that \tool achieves 38\% more branches, 57\% overhead reduction and 11.4$\times$ speed-up in bug exposure compared to \rev{a leading} state-of-the-art kernel concurrency fuzzer. 
Moreover, \tool discovers eight previously unknown concurrency-related bugs in the Linux kernel, \rev{six} of which have already been confirmed and fixed by developers.

\end{abstract}

\input{sections/01-Intro}

\input{sections/02-Background}

\input{sections/03-Motivation}

\input{sections/04-Design}

\input{sections/05-Implement}
\input{sections/06-Eval}
\input{sections/07-Case-study}
\input{sections/08-Limitations}

\input{sections/09-Perspectives}
\input{sections/Ethics}
\input{sections/OpenScience}
\input{sections/usenix26-ae}



\bibliographystyle{plain}
\bibliography{references}

\end{document}

%% file: packages.tex
\usepackage{usenix}
\usepackage[utf8]{inputenc}
\usepackage[english]{babel}

\usepackage{indentfirst} 

\usepackage{csquotes}

\usepackage{bookmark}
\usepackage{float}
\usepackage{inconsolata}

\usepackage{microtype} 
\usepackage{tabulary} 
\usepackage{hhline}

\usepackage{enumitem}

\usepackage[usenames,dvipsnames]{xcolor}

\usepackage{listings}
\usepackage[newfloat]{minted}

\usepackage{mathtools}

\usepackage{interval}
\usepackage[linesnumbered,ruled,vlined]{algorithm2e}

\usepackage{array}

\usepackage{color}
\usepackage[usenames,dvipsnames]{xcolor}
\usepackage{soul}
\usepackage{comment}

\usepackage{lipsum} 
\usepackage[inkscapelatex=false]{svg}
\usepackage{caption}
\usepackage{subfig}
\usepackage{booktabs}
\usepackage{multirow}
\usepackage{wasysym}
\usepackage{pifont}
\usepackage[linesnumbered,ruled,vlined]{algorithm2e}
\usepackage{amsmath}

%% file: macros.tex
\ifdraft
\newcommand{\editorial}[2]{{\color{#1}#2}}
\newenvironment{notes}{\color{blue}}{}
\else
\newcommand{\editorial}[2]{}
\excludecomment{notes}
\fi
\newcommand{\lijl}[1]{{\editorial{cyan}{Jialin: #1}}}
\newcommand{\djw}[1]{{\editorial{magenta}{Dylan: #1}}}

\newcommand{\jc}[1]{{\editorial{blue}{Jiacheng: #1}}}

\newcommand{\startrev}{}
\newcommand{\finishrev}{}
\newcommand{\rev}[1]{#1}

\newcommand{\scx}{\texttt{sched\_ext}\xspace}
\newcommand{\tool}{\textsc{Sect}\xspace}
\newcommand{\toolminus}{\textsc{Sect-}\xspace}
\newcommand{\toolrw}{\textsc{Sect}\_RW\xspace}
\newcommand{\toolpos}{\textsc{Sect}\_POS\xspace}
\newcommand{\toolpct}{\textsc{Sect}\_PCT\xspace}
\newcommand{\fuzzer}{\tool-fuzzer\xspace}
\newcommand{\executor}{\tool-executor\xspace}
\newcommand{\scheduler}{\textsc{Sect}-scheduler\xspace}
\newcommand{\xrw}{Priority Walk\xspace}
\newcommand{\schedext}{\texttt{sched\_ext}\xspace}

\newcommand{\circled}[1]{\textcircled{\raisebox{-0.9pt}{#1}}}

\newcommand{\ttt}[1]{\parbox{\linewidth}{
  \raggedright
  {\ttfamily
  #1
  }
}}

\let\svthefootnote\thefootnote
\newcommand\freefootnote[1]{%
  \let\thefootnote\relax%
  \footnotetext{#1}%
  \let\thefootnote\svthefootnote%
}

\definecolor{codekw}{RGB}{33,66,125} 
\definecolor{codetype}{RGB}{0,102,0}
\definecolor{codestring}{RGB}{163,21,21} 
\definecolor{codecomment}{RGB}{128,128,128}
\definecolor{codestring}{RGB}{163,21,21}
\definecolor{codeidentifier}{RGB}{0,0,0}
\definecolor{codenumber}{RGB}{128,128,128}
\definecolor{codeconstant}{RGB}{90, 90, 90}
\definecolor{codeadd}{rgb}{0.0, 0.5, 0.0}

\lstdefinestyle{CStyle}{
    language=C,
    aboveskip=1em, 
    belowskip=1em, 
    basicstyle=\ttfamily\scriptsize,  
    breaklines=true, 
    captionpos=b,          
    escapechar=@, 
    frame=none,
    keepspaces=true, 
    keywordstyle=\bfseries\color{codekw}, 
    float=false,
    commentstyle=\itshape\color{codecomment},         
    stringstyle=\color{codestring},              
    identifierstyle=\color{codeidentifier},             
    numbers=left, %
    numberstyle=\tiny\color{codenumber},
    numbersep=6pt, 
    showspaces=false,
    showstringspaces=false,  
    showtabs=false,
    stepnumber=1, 
    tabsize=4,
    xleftmargin=1em,
    xrightmargin=0em,
    emph={yield}, 
    emphstyle=\bfseries,
    morecomment=[f][\color{red}]{-},        
    morecomment=[f][\color{codeadd}]{+},      
    morekeywords=[2]{SCHED_EXT, TASK_RUNNING, NULL},
    keywordstyle=[2]\color{codeconstant},
}

%% file: sections/01-Intro.tex
\section{Introduction}
In 21st-century computing, the end of Moore's Law has led to a shift towards increasingly parallel and distributed software systems \cite{conc:1, conc:2, conc:3}.
Although concurrency is essential for achieving scalability and performance in modern OS kernels, 
it also introduces complex synchronization mechanisms that are notoriously error-prone \cite{conc:vuln}.
These bugs can lead to severe consequences due to the fundamental role of kernels in the software stack.
One prominent example is the Dirty COW vulnerability \cite{cve-2016-5195}, where a race condition exploited a flaw 
in the kernel's memory management.
This allowed an attacker to gain write access to sensitive memory mappings, facilitating local privilege escalation and potentially affecting millions of computing devices.
Thus, finding concurrency bugs is essential in ensuring a safe and reliable kernel.

Concurrency bugs are often more challenging to discover and identify
compared to their sequential counterparts \cite{conc:opensource}.
They are triggered with two important ingredients: 
execution of the buggy code path and observation of a particular thread interleaving.
Finding these bugs thus requires exploring not just the input space but also the thread interleaving space, which is both difficult to control and exponentially large in the number of concurrently executed instructions.

Historically, kernel concurrency testing has been constrained by a \textit{scheduling dilemma}. Conventional stress-testing \cite{stress:2, stress:CHESS, stress:java, stress:java2} and coverage-guided fuzzers \cite{actor, Syzkaller, kernelgpt, mock} typically treat the fairness- and latency-oriented kernel scheduler as an opaque component that must be overcome indirectly via exceptional workloads to manifest rare concurrency bugs.
To bypass this limitation, the state-of-the-art concurrency fuzzers have relied on external interposition via delay injection, modified hypervisors or invasive kernel patches \cite{Razzer, Conzzer, Snowboard, SegFuzz}.
While making notable progress over stress-testing by more directly affecting schedules from outside the kernel context, these approaches suffer from
the following challenges:

\textbf{C0: Limited scheduling control.} Many concurrency fuzzers either provide only weak control over thread scheduling by injecting delays~\cite{Krace,bechberger2025} or limited coarse-grained scheduling hooks~\cite{Razzer, Conzzer}.
With only limited, coarse-grained control, many concurrency bugs may be rendered unreachable during testing.
While trading completeness for performance is sometimes reasonable, these works fundamentally cannot enforce a higher degree of control, even in cases when performance is not an issue.

\textbf{C1: High overhead.} Traditional concurrency fuzzers with full control typically operate at the hypervisor level~\cite{SegFuzz, Snowboard}.
These hypervisor-specific techniques are extremely heavyweight and exhibit poor scalability. 
For example, Snowboard \cite{Snowboard} introduces a substantial performance overhead, slowing execution by more than $10\times$. Additionally, the limitations of the hypervisor's functionality restrict SegFuzz's support \cite{SegFuzz} for parallel fuzzing processes, resulting in a significant reduction in overall throughput during fuzzing (c.f. Section~\ref{sec:eval-cov}).

\textbf{\rev{C2: Portability and maintenance burden.}} OS kernels have been under continuous development, with frequent updates and ongoing evolution.  In addition to custom hypervisors, existing fuzzers usually require extensive customization tailored to specific kernel versions \cite{SegFuzz, Krace, Razzer}. For instance, Krace \cite{Krace} involves invasive kernel changes, introducing a patch of over 10k lines of code across 105 files. Such tight coupling substantially increases the cost and effort required to port the fuzzer to newer kernel versions, and may also reduce its effectiveness. As a result, the long-term usability and maintainability of such approaches are often compromised.

\textbf{C3: Restricted extensibility.} Scheduling algorithms in existing concurrency fuzzers are often bespoke and tightly integrated with critical components of the underlying implementation.
Extending or duplicating them in a new codebase is typically extremely challenging and demands deep domain expertise.
As a result, traditional fuzzers that rely on default scheduling behavior are often still used to attempt to find concurrency bugs in practice. 

\input{figures/sota}

Table \ref{tab:status} summarizes existing kernel concurrency fuzzing approaches and their limitations. 
In this work, we rethink kernel concurrency testing by shifting from external intervention to native exploration. 
We propose a \emph{scheduler-as-an-explorer} paradigm: rather than treating the scheduler as an opaque entity to be perturbed externally, we elevate it into  a first-class mechanism for systematic interleaving exploration. We propose \tool (\textbf{S}ched-\textbf{E}xt \textbf{C}oncurrency \textbf{T}ester), the first \emph{kernel-native} concurrency fuzzing framework. \tool introduces a novel CCT scheduler based on temporal isolation beyond traditional performance-driven scheduling and embeds interleaving exploration directly into the kernel scheduling path via eBPF. 
Therefore, \tool enables fine-grained and lightweight control over thread interleavings without requiring hypervisor modifications or invasive kernel patches. 
In addition, \tool incorporates programmable and testing-oriented scheduling algorithms and supports their modular extension, enabling future research in concurrency exploration strategies.
To further increase exploration granularity, \tool provides a preemption-safe instrumentation mechanism for injecting scheduling points at critical kernel events. Finally, \tool integrates with Syzkaller to facilitate a two-phase fuzzing workflow, jointly exploring sequential and concurrent behaviors.

We apply \tool to recent Linux kernel versions and conduct extensive evaluation.
Our results demonstrate that the \emph{scheduler-as-an-explorer} paradigm is highly effective: \tool achieves 38\% higher branch coverage than the state-of-the-art concurrency fuzzer SegFuzz. \tool also accelerates bug reproduction by 11.4$\times$ on known concurrency bugs and provides a 57\% reduction in overhead compared to SegFuzz. Additionally, \tool discovers eight previously unknown concurrency-related bugs in recent Linux kernels, \rev{six} of which have already been confirmed and fixed by developers.

In summary, our work makes the following contributions:
 \begin{itemize}
\item We identify the \emph{scheduling dilemma} as a fundamental limitation of existing kernel CCT techniques and propose a new \emph{scheduler-as-an-explorer} paradigm, mitigating the challenges {\bf C0}, {\bf C1}, {\bf C2} and {\bf C3}. 
\item We design and implement \tool, the first kernel-native concurrency fuzzing framework that features the temporal isolation scheduling with programmable policies, the safe preemption injection mechanism, and the two-phase input-and-concurrency fuzzing workflow.
\item We evaluate \tool on recent Linux kernel versions, demonstrating substantial improvements in coverage, efficiency, and bug discovery effectiveness.
\item We make \tool available and open-source, lowering the entry barrier for research in CCT and concurrency fuzzing for kernel code.
 \end{itemize}

%% file: figures/sota.tex
\begin{table}[t!]
\scriptsize
\centering
\caption{State-of-the-art kernel concurrency fuzzers.
}
\label{tab:status}
\setlength{\tabcolsep}{1mm}{
\begin{tabular}{@{}lccccc@{}}
\toprule
\textbf{Tool} &
  \textbf{\begin{tabular}[c]{@{}c@{}}Granular\\ Control\end{tabular}} &
  \textbf{\begin{tabular}[c]{@{}c@{}}Low-\\ Overhead\end{tabular}} &
  \textbf{\begin{tabular}[c]{@{}c@{}}Hypervisor\\ Agnostic\end{tabular}} &
  \textbf{\begin{tabular}[c]{@{}c@{}}Forward\\ Compatible\end{tabular}} &
  \textbf{\begin{tabular}[c]{@{}c@{}}Highly\\ Extensible\end{tabular}} \\ \midrule
Razzer \cite{Razzer}       & \textcolor{magenta}{\ding{55}} & \textcolor{magenta}{\ding{55}} & \textcolor{magenta}{\ding{55}} & \textcolor{green}{\ding{51}}  & \textcolor{magenta}{\ding{55}} \\
Snowboard \cite{Snowboard} & \textcolor{green}{\ding{51}} & \textcolor{magenta}{\ding{55}} & \textcolor{magenta}{\ding{55}}   & - & \textcolor{magenta}{\ding{55}} \\
SKI~\cite{SKI} & \textcolor{green}{\ding{51}} & \textcolor{magenta}{\ding{55}} & \textcolor{magenta}{\ding{55}}   & \textcolor{magenta}{\ding{55}}        & \textcolor{magenta}{\ding{55}} \\
KRACE \cite{Krace}         & \textcolor{magenta}{\ding{55}}   & \textcolor{magenta}{\ding{55}} & \textcolor{green}{\ding{51}} & \textcolor{magenta}{\ding{55}}       & \textcolor{magenta}{\ding{55}}    \\
SegFuzz \cite{SegFuzz}     & \textcolor{green}{\ding{51}} & \textcolor{magenta}{\ding{55}} & \textcolor{magenta}{\ding{55}}   & \textcolor{magenta}{\ding{55}}        & \textcolor{magenta}{\ding{55}} \\ \midrule
\textbf{\tool}             & \textcolor{green}{\ding{51}}   & \textcolor{green}{\ding{51}}   & \textcolor{green}{\ding{51}}   & \textcolor{green}{\ding{51}} &  \textcolor{green}{\ding{51}} \\ \bottomrule
\end{tabular}
}
\end{table}

%% file: sections/02-Background.tex
\section{Background and Related Work}

\subsection{Kernel Fuzzing}
Fuzzing has become the de facto standard technique for dynamically uncovering OS kernel vulnerabilities \cite{aflplusplus, aflgo, stateafl, miner, park2020fuzzing, mopt}.
Fuzzers~\cite{statefuzz, syz:2, actor, syz:6, syzvegas, healer}  explore kernel states in a biased randomized search by prioritizing test inputs that cover new branches.
For instance, Syzkaller, the state-of-the-art kernel fuzzer, \cite{Syzkaller} generates system call (syscall) programs as inputs for the Linux kernel, primarily meant for sequential execution; it has identified over 5,000 bugs in the upstream Linux kernel but only roughly $10\%$ are concurrency-related issues.
The approach taken by these conventional, sequential-oriented fuzzers enforces \emph{no control} over the thread interleavings, relying entirely on the native OS scheduler; this greatly restricts their ability to explore the scheduling space to discover rare concurrency bugs. 

To fill this gap, several concurrency-oriented fuzzers have been proposed to handle exploration of concurrency space.
Krace \cite{Krace} introduces an alias coverage that is implemented  to guide fuzzing towards new interleavings.
However, Krace adopts a lightweight, but weak form of scheduling control by injecting random delays at memory access points.
Delay injection alone, however, is likely to miss many uncommon thread interleavings during fuzzing~\cite{Conzzer} (\textbf{C0}).
Moreover, Krace requires substantial modifications of the Linux kernel, making it difficult to maintain as the kernel evolves over time (\textbf{C2}).

Razzer \cite{Razzer} modifies the hypervisor to include hypercalls that, when called by a kernel thread, would install hardware breakpoints on the vCPU for scheduling.
These hardware breakpoints are installed at memory reads and writes, which are pre-determined by a heuristic-driven static analysis.
To trigger a race, Razzer does a single step on the specified vCPU before resuming execution on all vCPUs. 
However, Razzer can still only enforce control at a small number of hardware breakpoints (typically four on current generation Intel processors), limiting its ability to explore schedules involving more than four events (\textbf{C0}). 

SegFuzz \cite{SegFuzz}, which builds on top of the Razzer architecture, improves on Razzer's limitation by replacing hardware breakpoints at runtime, enforcing a fully controlled schedule sent by the fuzzer process via hypercalls.
Nevertheless, we observe that SegFuzz, constrained by its reliance on the hypervisor and virtualized hardware: it is limited to a single fuzzing process per virtual machine and thus does not scale well to larger fuzzing campaigns (\textbf{C1}).

SKI \cite{SKI} and Snowboard \cite{Snowboard} use a similar method of sending hypercalls to suspend and resume the execution of vCPUs according to a schedule.
However, instead of hardware breakpoints, they make use of the ability to set a thread's CPU affinity to pin each thread to its own vCPU.
While this avoids the limited parallelism of SegFuzz, it introduces high latency for fine-grained context switches, as it does not leverage hardware features to achieve fast preemption (\textbf{C0}).



\subsection{Controlled Concurrency Testing }
A strongly related research to concurrency fuzzing is Controlled Concurrency Testing (CCT).
These works focus on exploring the space of interleavings of a \emph{fixed input}, rather than simultaneously searching the space of inputs and interleavings as in concurrency fuzzing.
Many systematic~\cite{musuvathi2008finding,contextbound2005} and randomized~\cite{wolff2024greybox, sen2007effective} CCT scheduling algorithms have been developed which can effectively navigate the interleaving space.
In practice, randomized algorithms tend to outperform their systematic counterparts at scale~\cite{thomson2016concurrency}, and some of these randomized algorithms even provide probabilistic guarantees for finding certain classes of bugs~\cite{PCT, yuan2018partial, zhao2025surw}.
These algorithms have seen deployments in large-scale, industrial settings at companies like Amazon~\cite{bornholt2021using} and Microsoft \cite{coyote} for testing user-space programs.
However, with the notable exception of SKI~\cite{SKI}, which implements only the PCT~\cite{PCT} algorithm, CCT scheduling algorithms remain largely unevaluated on OS-kernel software, which represents one of the most important target concurrent applications.
As such an explicit goal of this work is to make these algorithms available in an extensible framework for kernel code (\textbf{C3}).

\subsection{eBPF Technology}
The Extended Berkeley Packet Filter (eBPF) enables dynamic in-kernel customization at runtime without modifying the Linux kernel code.
It allows developers to load and run restricted programs in a safe, sandboxed, in-kernel virtual machine (VM) \cite{mohamed2023under, dejaeghere2023comparing}.
These programs attach to specific kernel hook points and interact with the kernel through vetted helper calls.
By allowing developers to inject custom logic into the kernel at runtime, eBPF provides a flexible and low overhead means to monitor and manipulate kernel behavior. 

One notable application of eBPF is the \scx ~\cite{daniel2024sched, sched} scheduler class, which has been officially merged into the mainline Linux kernel since version 6.12.
Traditionally, process scheduling in Linux has been limited to a fixed set of in-kernel policies, such as the Completely Fair Scheduler (CFS)~\cite{CFS}.
\texttt{sched\_ext} decouples scheduling policy from the core kernel scheduling mechanism.
It exposes user-defined scheduling logic through public interfaces \cite{yasukata2024developing}, and allows developers to implement customized scheduling policies through eBPF programs.
Only tasks explicitly set with the \texttt{SCHED\_EXT} scheduling class are affected by programmable policies.
While \texttt{sched\_ext} is originally designed for domain-specific schedulers for performance optimization, our novelty is to leverage this feature to support kernel fuzzing for CCT.

%% file: sections/03-Motivation.tex
\section{Motivation}

Many concurrency bugs manifest only under specific instruction orderings across multiple threads.
However, kernel schedulers are primarily designed to optimize fairness and performance rather than to expose such rare thread interleavings, making the natural occurrence of concurrency bugs unlikely.
Thus, the ability to manipulate scheduling behaviour  can be a powerful strategy to find and reproduce concurrency bugs.

\sloppy

\subsection{Motivating Example}

\input{figures/jfs-partial}

To show the need for a CCT tool for the Linux kernel, we examine a concurrency bug that we found in the \texttt{fs/jfs} module that results in a null pointer dereference. 
It is closely related to a recent bug which had been reported as fixed \cite{d6c1b3}.
The previous issue was a use-after-free (UAF) due to missing synchronization in the \texttt{jfs\_ioc\_trim} function that allowed it to interleave execution with \texttt{dbUnmount}.
We show both functions in Listing~\ref{fig:jfs-partial}, along with the full patch we provided to developers.
The UAF was triggered when \texttt{dbUnmount} freed \texttt{bmap} in one thread while \texttt{jfs\_ioc\_trim} still had a reference to it in another thread.
It was resolved by adding a read-write semaphore such that the free cannot interleave between obtaining the pointer and dereferencing it in \texttt{jfs\_ioc\_trim}.

Unfortunately, the accepted original patch omitted the null check at line 16, and was thus still susceptible to a null pointer dereference.
This issue was not caught by kernel developers but was found by \tool.
In particular, \texttt{JFS\_SBI(ipbmap->i\_sb)->bmap} is set to null in \texttt{dbUnmount} after it is freed.
Thus, while the UAF bug can no longer occur, if the critical section in \texttt{dbUnmount} executes first, the resulting pointer will be null instead of the old address.
As the pointer is dereferenced in \texttt{jfs\_ioc\_trim}, this leads to a null pointer dereference under some interleavings. 
We observed that this bug cannot be reproduced by the default OS scheduler in 10,000 executions of a syscall sequence. In contrast, 
\tool consistently reproduces the issue in fewer than ten executions.
This case highlights the need for more effective concurrency testing for kernel code.

\subsection{From Scheduling Black-Box to Scheduler-as-an-Explorer}
In principle, kernel concurrency testing could be realized by implementing a dedicated scheduler class directly within the Linux kernel.
Such an approach would offer stronger control over scheduling decisions, but it comes at the cost of invasive kernel modifications, compromised safety, and limited deployability. More importantly, tightly coupling a concurrency-testing scheduler with the kernel core contradicts our goal of enabling flexible and extensible exploration of scheduling behaviors (\textbf{C3}).
To realize the \emph{scheduler-as-an-explorer} paradigm while preserving kernel safety and practicality, we build our CCT scheduler on top of the \texttt{sched\_ext} infrastructure and eBPF.
This design choice enables systematic control over thread interleavings without modifying the kernel core or relying on external enforcement mechanisms.

\noindent\textbf{Safety and Reliability.}
The \scx-based schedulers are validated and sandboxed by the eBPF compile-time verification and runtime checks, which enforce strict safety guarantees.
These mechanisms eliminate a broad class of implementation errors that could otherwise lead to kernel crashes.
Avoiding these failures is particularly important for kernel concurrency testing, as unintended crashes can prematurely terminate testing campaigns.

\noindent\textbf{Deployability and Portability.}
A \scx scheduler can be loaded, attached, and replaced at runtime without kernel rebuild or system reboot.
The eBPF and \scx APIs have remained stable over recent years, providing a consistent interface for scheduler development \cite{sched}.
In addition, the presence of BPF Type Format \cite{btf}, libbpf \cite{libbpf}, and the eBPF compiler enables scheduler portability without rebuilds across kernel versions. 
\rev{While we cannot guarantee complete forward compatibility, we believe that} our approach significantly reduces maintenance effort and improves the long-term practicality of kernel concurrency testing (\textbf{C2}).

%% file: figures/jfs-partial.tex
\begin{lstlisting}[
style=Cstyle, 
caption={A patch completing a previous partial fix  in \texttt{jfs} filesystem.},
label={fig:jfs-partial}]
// fs/jfs/jfs_discard.c
int dbUnmount(struct inode *ipbmap, ...) {
    // ...    
    kfree(bmp);
    JFS_SBI(ipbmap->i_sb)->bmap = NULL;
}

int jfs_ioc_trim(...) {
-   struct bmap *bmp = JFS_SBI(ip->i_sb)->bmap;
+   struct bmap *bmp;
    // ...
+   down_read(&sb->s_umount);
+   bmp = JFS_SBI(ip->i_sb)->bmap;

-   if (minlen > bmp->db_agsize ||
+   if (bmp == NULL ||
+       minlen > bmp->db_agsize ||
 	    start >= bmp->db_mapsize ||
-       range->len < sb->s_blocksize ||
+       range->len < sb->s_blocksize) {
+       up_read(&sb->s_umount);
 		return -EINVAL;
+   }
    // ...
+   up_read(&sb->s_umount);
 	range->len = trimmed << sb->s_blocksize_bits;
 	return 0;
}
\end{lstlisting}

%% file: sections/04-Design.tex
\section{Design of \tool}

\input{figures/overview}

\tool alternates between two-phases of fuzzing: sequential input exploration and concurrent schedule fuzzing.
The first phase is sequential fuzzing, which aims to explore the input space with traditional generation and mutation strategies, maximizing code coverage.
The second phase, concurrency fuzzing, constitutes the core of \tool and focuses on exploring the thread interleaving space.

\autoref{design:overview} illustrates the design components and the overall workflow of \tool.
To increase exploration granularity in this phase, \tool optionally instruments the kernel under test with additional preemption-safe scheduling points at compile time (\circled{1}).
The \tool scheduler can be loaded once at test-time (\circled{2}) without needing to reboot or recompile the kernel.
During the concurrency fuzzing, \tool employs a concurrency-aware mutation strategy (\circled{3}) to generate multi-threaded syscall programs.
As the core component of \tool, we propose a novel CCT scheduler based on temporal isolation scheduling to control the concurrent execution via eBPF (\circled{4}).
During the execution of a multi-threaded test input (\circled{5}) in the concurrency fuzzing phase, \tool delegates the scheduling control of target tasks to the CCT scheduler.
When a scheduling point (e.g. a CPU yield) is reached, \tool enforces a schedule decision and selects the next thread to run from the set of runnable candidates according to specific CCT scheduling strategies.
\tool repeats this scheduling process until all concurrent threads are finished.
In doing so, \tool can effectively serialize and control (\circled{6}) multi-threaded executions into temporal isolation, thereby increasing the likelihood of triggering bugs that manifest due to rare interleavings.
\tool always monitors whether the kernel under test crashes or terminates with any errors, in addition to collecting code coverage feedback (\circled{7}) to drive exploration of the input space. 

Through steps \circled{1} - \circled{7}, there are four main components in the design of \tool,  as shown in blue in \autoref{design:overview}:
the CCT scheduler, the programmable scheduling strategies, the preemption-safe instrumentation, and the fuzzing loop.
In the following section, we provide a detailed description of each design component.




\input{figures/lifecycle}

\subsection{Temporal Isolation Scheduling}
\label{sec:scheduler}

We define \textit{temporal isolation} as a scheduling paradigm in which (a) tasks can be partitioned into domains, with each domain's scheduling decisions remaining isolated from the other and (b) only one task from each domain is running at a time, serializing the execution.
The goal of the \tool scheduler is to explore the interleaving of events on different kernel tasks to expose concurrency vulnerabilities via temporal isolation scheduling.
In this section, we present the core logic of the \tool scheduler and how it realizes temporal isolation scheduling using eBPF.

\textbf{Task Lifecycle}
We begin by briefly reviewing the task lifecycle model of the \texttt{sched\_ext} scheduling class illustrated in \autoref{design:lifecycle}. 
In kernel terminology, both threads and processes are uniformly represented and scheduled as tasks, and we use these terms interchangeably.
Each task is created in the \texttt{init\_task} state and enqueued in a run queue and later dispatched to a CPU with an assigned time budget.
A task transitions to a quiescent state either by exhausting its time slice or voluntarily yielding the CPU due to blocking operations \cite{sched}.

\textbf{Temporal Isolation Serialization.} 
Rather than treating scheduling merely as a performance-oriented resource allocation mechanism, \tool adopts a \emph{scheduler-as-an-explorer} perspective, in which the scheduler actively drives the exploration of concurrent behaviors.
To enable this exploration, we enforce \emph{temporal isolation serialization} among tasks under test.
Concretely, the \tool scheduler serializes task execution by dispatching a runnable task only after ensuring that all other eligible tasks are either quiescent or safely re-enqueued.
Once a task is dispatched, the scheduler will not dispatch any additional tasks (even if idle cores are available) until the running task either blocks or yields its CPU.
The scheduler iteratively repeats this process until all tasks involved in the test have terminated.

Under this temporal isolation serialization paradigm, concurrent execution proceeds as a sequence of kernel events rather than an imprecise, uncontrolled set of parallel events.
As a result, temporal isolation serves as an exploration primitive that grants programmable control over concurrency, enabling the scheduler to systematically expose memory races
and synchronization interactions that would otherwise remain implicit under uncontrolled parallel execution.

\input{figures/algo}

\textbf{Partitioning of Scheduling Domains.} While the temporal isolation serialization facilitates controlled exploration within a single test, enforcing it globally across the entire system would result in severe underutilization of system resources.
To support high-throughput testing, \tool partitions tasks from concurrent test instances into isolated scheduling domains.
Specifically, the scheduler of \tool
identifies these domains by inspecting attributes within the 
task structure and leverages these attributes (e.g., command name or group identifiers), to cluster tasks belonging to the same test instance into a unified domain.
At each scheduling event, the scheduler first resolves the task's domain membership and then enforces the temporal isolation serialization
discipline strictly within that domain.
This mechanism allows multiple test instances to execute in parallel without interference. In this way, \tool decouples the scheduling decisions for each domain, ensuring that the controlled scheduling are applied locally without impeding parallel execution.

Algorithm \ref{alg:controlled-scheduling} presents the core scheduling procedure of \tool scheduler.
When the scheduler is invoked, either due to timer interrupt or the running task transitioning to the quiescent state, the algorithm locates the corresponding scheduling domain (line 3), removes the current task if it is exiting (line 6) or disables the task from scheduling if it is blocked (line 4).
Otherwise, the weight of each task in the domain (line 10) are updated based on CCT strategies that is discussed in Section~\ref{sec:sched-algo}.
Finally, the scheduler selects an enabled task from the domain with the highest weight and dispatches the task to run.
This architecture enables flexible task scheduling to support diverse CCT strategies, addressing \textbf{C3}:
only the weight-assignment function needs to be defined, without requiring modifications to the overall scheduling lifecycle.

\textbf{Shadow Scheduling Metadata.}
While \tool leverages the \texttt{sched\_ext} feature to interact with the native kernel, this infrastructure is designed for uncontrolled scheduling and optimized for latency and fairness, rather than exploration. This leads to several impedance mismatches for CCT, which \tool overcomes.
For example, while the Linux scheduling infrastructure provides a native priority queue interface for task dispatch \cite{sched_ext:article}, it prohibits modifying task attributes (e.g., priority) once a task has been enqueued.
This restriction fundamentally limits its applicability for CCT, as many scheduling strategies (such as PCT~\cite{PCT} and POS~\cite{yuan2018partial}) require dynamic, event-driven priority updates based on runtime states. 
This dynamism is crucial for reacting to program behaviors as they unfold. For example, when a running task is observed to race with a paused task, the scheduler may need to elevate the priority of the paused task to realize a specific interleaving~\cite{yuan2018partial}. 
Therefore, the native dispatch queue is insufficient to support exploration-oriented CCT workloads, directly relating to \textbf{C1}.

To overcome this limitation, \tool bypasses the native kernel data structures and instead maintains a shadow scheduling metadata layer  implemented using eBPF maps \cite{ebpf_maps}.
This shadow structure stores per-task metadata, including dynamically adjustable priority values, scheduling eligibility, domain membership, and other runtime attributes required by CCT scheduling strategies. 
In addition, priority computation and updates are delegated to the scheduling strategies themselves rather than being embedded in the dispatch mechanism.
As later discussed in Section~\ref{sec:sched-algo}, this \rev{decoupling} of scheduling logic from dispatch logic enables different CCT strategies to define, refine, and evolve their own policies with minimal overhead, addressing \textbf{C3}.

\subsection{Programmable Scheduling Strategies}
\label{sec:sched-algo}

A scheduling strategy in \tool represents a specific algorithm by defining a weight-assignment function invoked at each scheduling point.
\tool exposes this extensible and programmable scheduling interface for diverse algorithms which promote exploration of the interleaving space, rather than latency or fairness.
By interacting with the shadow scheduling metadata, these algorithms can react to dynamic program states in real-time, overcoming the inherent rigidity of native kernel scheduling interfaces. 
To demonstrate the versatility of \tool, we have implemented four representative exploration-oriented algorithms: Random Walk, Random Priority, POS~\cite{yuan2018partial}, PCT~\cite{PCT}.
\tool is also extensible to support other more sophisticated scheduling strategies.

\textbf{Random Walk.}
Random Walk (RW) is our most naive scheduling policy.
At each scheduling point, we select the next task to run from the pool of enqueued tasks uniformly at random. The algorithm assigns the same weight to all runnable tasks.
Notably, RW provides probabilistic fairness --- tasks are unlikely to be delayed for an extended period of time.
It also results in many preemptive context switches, as each scheduling decision has a $\frac{N-1}{N}$ probability of switching tasks if there are $N$ enqueued tasks available to run.

\textbf{Random Priority.}
For the Random Priority (RP) policy, at each scheduling point, we assign the most recently executed task a priority.
The algorithm then picks the runnable task with the highest weight to execute next.
This approach differs from RW in that the weights of most of the tasks persist across scheduling points.
As a result, RP is prone to starvation, in that a task with low weight will likely not be selected by the algorithm, and thus stick with its low priority for subsequent scheduling decisions.

\textbf{Partial Order Sampling.}
In Partial Order Sampling (POS), each task is assigned a random weight, and the weight of the most recently executed task is reset at each scheduling point, similar to the RW policy.
POS further re-assigns the weight of any task whose event interferes with the most recent event of the previously scheduled task~\cite{yuan2018partial}.
Here, interference refers to accessing the same heap object, with at least one access being destructive (e.g., a \texttt{load} and \texttt{store} instruction accessing the same address).
This adjustment enables POS to select different orderings of interfering events with equal probability. Consequently, POS is the only semantically aware concurrency testing algorithm in \tool, requiring source code information to compute interference. Similar to RP, POS can also suffer from starvation.

\textbf{PCT.} Probabilistic Concurrency Testing (PCT) \cite{PCT} is a well-known algorithm developed for concurrency testing that provides a lower bound on the probability of detecting \emph{low depth} concurrency bugs -- parameterized by $d$.
It is an extremely \emph{unfair} algorithm: It only enforces $d-1$ context switches in an entire execution of thousands, or even millions of events, where $d$ is often recommended to be set as low as three~\cite{thomson2016concurrency}.
In PCT, $d$ intervals are sampled uniformly from possible locations in the program execution.
Tasks are assigned static weights at the beginning of execution.
At each of the $d$ intervals, the task with the highest weight is de-prioritized, giving the second highest weighted task primacy.
To compute the interval sizes, PCT requires an estimate of the total number of events in a given execution, usually approximated by counting the events of a ``trial run''.


\subsection{Scheduling Point Injection}
\label{sec:inst}

Under temporal isolation scheduling, once a task is dispatched by \tool, it continues execution until it exhausts its assigned time slice or voluntarily yields its CPU, at which point the scheduler re-gains control \cite{sched}. 
Although kernels already contain some existing scheduling points, such as blocking operations and slice expires, these points are primarily designed to ensure fairness and responsiveness rather than to expose concurrency bugs.
Indeed, many kernel concurrency vulnerabilities arise from subtle re-orderings around shared-memory accesses and synchronization operations.
Consequently, relying solely on existing native scheduling points can leave a large portion of the concurrency-relevant interleaving space unexplored.
To bridge this semantic gap, \tool introduces additional preemptions comprehensively throughout the kernel, elevating these events to explicit scheduling points.
As a result, it aligns scheduling control with the semantic locations more likely to trigger concurrency bugs.

\textbf{Identifying Preemption.}
In principle, any operation that allows tasks to synchronize or otherwise communicate with each other could lead to buggy behavior if reordered.
Therefore \tool targets two classes of runtime events that mediate the vast majority of inter-thread interactions: shared-memory accesses and explicit synchronization operations.
Specifically, we identify and instrument memory access operations at the LLVM IR level~\cite{llvm} (such as \texttt{load} and \texttt{store} instructions) which are not marked as thread-local by the compiler.
In addition to shared-memory accesses, we instrument a set of high-level synchronization primitives which we manually identified in the kernel source code, including spin locks, semaphores, RCU synchronization, and other widely used kernel concurrency constructs.
By injecting scheduling points \emph{only} at semantically motivated locations, \tool avoids unnecessary overhead of context switching at points which do not involve communication between tasks and thus cannot trigger additional concurrency bugs.
While we believe this set of additional preemptions to be highly comprehensive and sufficient for finding most concurrency bugs in practice, we discuss the completeness limitations of \tool in Section \ref{sec:limit}.
We note that potential incompleteness arising from missing synchronization points can be mitigated by extending the instrumentation to inject scheduling points at such locations.

To control runtime overhead introduced by the additional instrumentation of the kernel, \tool supports \rev{configurable} probabilistic triggering of the scheduler at injected points.
By default, the scheduler is invoked with a 10\% probability at each instrumented location during fuzzing.
This design \rev{exposes} fine-grained control over the trade-off between exploration depth and execution throughput without sacrificing probabilistic completeness (Section \ref{sec:limit}), and can be readily adapted to different scheduling algorithms or kernel targets.

\textbf{\rev{Preemption Safety.}} After identifying a set of candidate scheduling points, \tool instruments these locations with an explicit call to the scheduler of \tool. 
However, scheduling points in kernel code must adhere to the implicit invariants governing where voluntary context switches are permitted.
For example, preempting execution within interrupt contexts, non-preemptible regions, or critical sections protected by spin locks may lead to kernel instability or deadlocks. 
To preserve execution fidelity, \tool first performs a runtime preemption admissibility check that evaluates the safety of a potential scheduling point prior to intervention.
Instead of attempting to formalize all non-preemptible regions, \tool leverages kernel native synchronization assertions, such as predicates derived from \texttt{might\_sleep()} in the Linux kernel. These predicates encapsulate developer-defined safety invariants, indicating whether the current context permits re-scheduling.
Once a candidate point satisfies the admissibility check \rev{at \emph{runtime}}, \tool triggers a voluntary CPU yield. This yield serves as the entry point for scheduler intervention, allowing \tool to suspend the current task and perform a context switch according to the active exploration strategy.

\startrev Notably, while generally conservative, the admissibility checks are not a \emph{guarantee} of safety. 
However, we find they work extremely well in practice to avoid false-positive bug reports due to unsafe preemptions (discussed in Section \ref{sec:limit}).
As demonstrated in our evaluation of an ablative variant of \tool without additional scheduling point instrumentation (Section~\ref{sec:eval}), this mechanism generally maintains the invariants of the kernel under test while enabling more fine-grained schedule exploration \finishrev.

\subsection{Fuzzing Loop}

To enable continuous test generation and execution, \tool incorporates the CCT-oriented scheduler into the fuzzing process. \tool alternates between two phases: sequential fuzzing and concurrency fuzzing, each of which is described below.

\textbf{Sequential Fuzzing.} The primary goal of sequential fuzzing is to explore a broader range of execution paths in the kernel.
In this phase, \tool delegates completely to Syzkallers' existing input-fuzzing strategy \cite{Syzkaller}, which we summarize below.
Syzkaller relies on syscall specifications written in Syzlang \cite{syzlang} to guide input generation and mutation towards valid syscall sequences. 
During test generation, Syzkaller selects syscalls at random from candidates,  
appends them to the end of the current sequence, and instantiates their parameters.
The resulting sequences are then executed successively in their original order by the kernel under test.
If a sequence triggers previously unexplored code branches, it is considered promising and will be preserved in the corpus for further mutation. 
During test mutation, Syzkaller picks up a seed from the corpus and performs one of mutation strategies in an attempt to uncover new coverage. Syzkaller employs traditional mutation strategies, such as inserting additional syscalls, removing existing syscalls, and modifying the values of the parameters of syscalls.
While for now, \tool merely leverages the default exploration strategy of Syzkaller, in the future, \tool can integrate more heavy-weight input generation techniques~\cite{syzvegas, kernelgpt, Snowboard}.
However, because sequential input fuzzing is orthogonal to the core contributions of \tool, we leave such enhancement to future work.
\input{figures/algo_mutation}

\textbf{Concurrency Fuzzing.} 
Existing fuzzers such as Syzkaller usually use ad-hoc approaches to increase the chances of interesting concurrent behaviors without controlling the schedule.
\tool proposes a custom concurrency mutation strategy tailored to the controlled temporal isolation scheduling provided by the \tool scheduler. 
Algorithm \ref{alg:mutation} presents the overall procedure of the mutation strategy.
The input to this algorithm is test cases consisting of syscall sequences $p$, generated via the sequential fuzzing phase.
For a given seed input $p$, \tool begins by randomly picking up $N$ syscalls as $S$ from the sequence to execute concurrently. 
The subset $S$ represents the targeted syscalls for scheduling control.
\tool then replicates each syscall in $p$ to the new sequence $P$.
During this process, if a syscall belongs to the selected subset $S$, it is also duplicated in the new sequence.
This duplication strategy introduces opportunities for concurrent behavior while maintaining the structural and semantic correctness of the original sequence. 

To facilitate concurrent exploration, \tool leverages an existing annotation mechanism provided by Syzkaller to designate specific syscalls in $P$ as asynchronous. 
At runtime, a specialized executor derived from Syzkaller interprets these annotations and instantiates each annotated syscall with its own dedicated thread.
These threads are bound to the  \texttt{SCHED\_EXT} scheduling class \cite{sched}, effectively delegating their execution control to \tool scheduler.
In contrast, unannotated syscalls run under the native scheduler, incurring no additional scheduling overhead.
To establish a consistent starting state for interleaving exploration, the executor injects a synchronization barrier, aligning the entry points of all participating
tasks before initiating controlled scheduling. 
Once the barrier is released, the temporal isolation scheduling described in Section~\ref{sec:scheduler} begins and explores the interleaving space of targeted syscalls for multiple iterations.

We exemplify the mutation and execution process in \autoref{design:overview}.
For a given program consisting of syscalls $\{S_1, S_2, S_3\}$, \tool may mutate it to mark $S_1$ and $S_3$ as asynchronous (\circled{3}).
Each of these syscalls has some preemption events  $\{e_1, e_2, e_3\}$ and $\{e_4, e_5, e_6\}$ which are separated by scheduling points injected in Section~\ref{sec:inst}.
During the execution phase (\circled{5}),  \tool cross-interleaves the events of the asynchronous syscalls $S_1$ and $S_3$ (\circled{6}) to expose potential race conditions, while the execution of other syscalls remains unaffected.

%% file: figures/overview.tex
\begin{figure*}[h]
\centerline{\includegraphics[width=\textwidth]{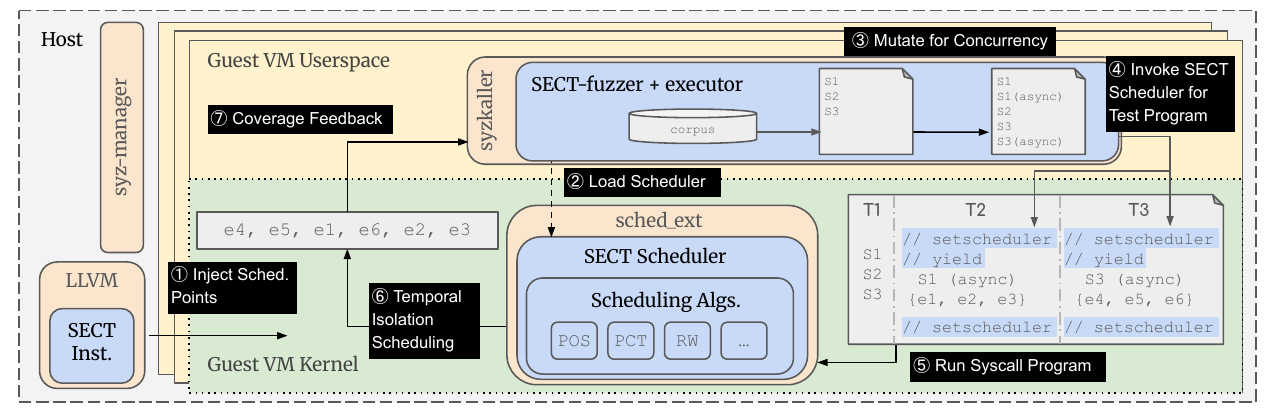}}
\caption{Overview of \tool (components of \tool in blue).}
\label{design:overview}
\end{figure*}

%% file: figures/lifecycle.tex
\begin{figure}[htbp]
\centerline{\includegraphics[width=\columnwidth]{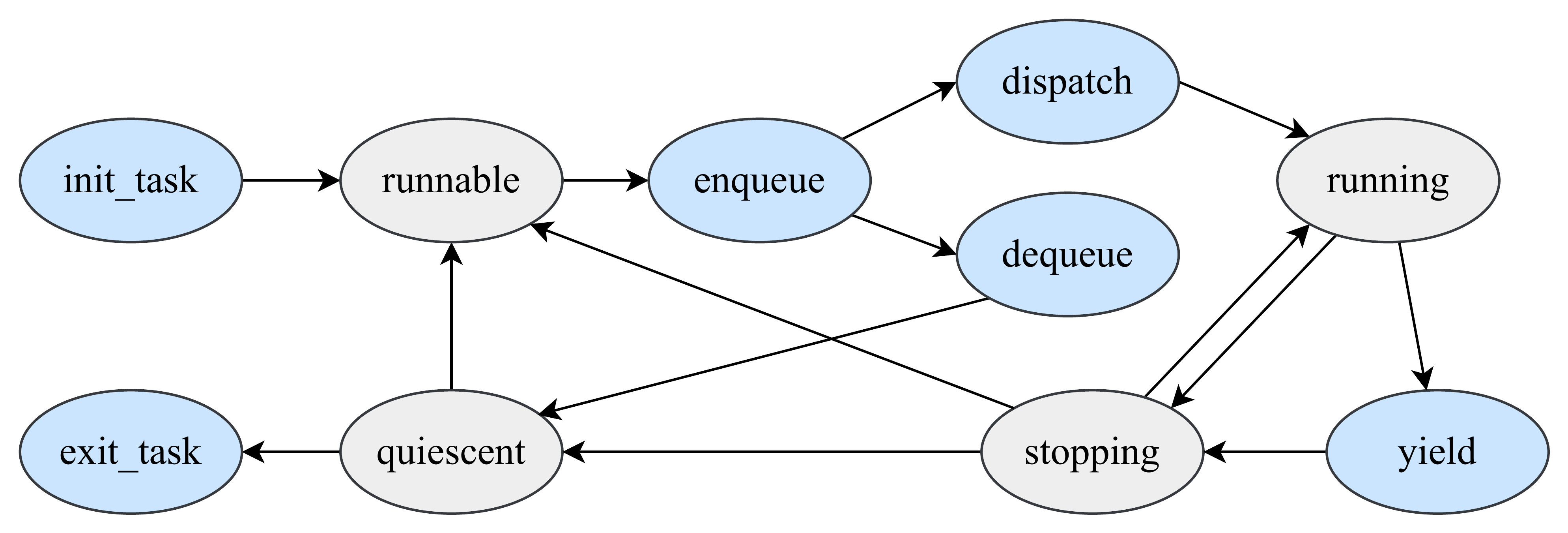}}
\caption{Task lifecycle within the \texttt{sched\_ext} infrastructure. Grey ovals represent the states that a task can be in, and blue ovals represent actions that a task can undergo.
}
\label{design:lifecycle}
\end{figure}

%% file: figures/algo.tex
\begin{algorithm}[htb]
\small
\caption{Temporal Isolation Scheduling}
\label{alg:controlled-scheduling}
\DontPrintSemicolon
\SetKwInput{Input}{Input}
\SetKwInput{Output}{Output}

\Input{set of active tasks $T$}
\Input{current task $t$}
\Output{serialized execution trace $S$}

$\textit{initialize\_scheduling\_algorithm}()$\;

\While{$|T| > 0$}{
    $d \gets \textit{get\_tasks\_in\_domain}(t)$\;
    \If{$\textit{is\_sleeping}(t)$}{
        $t.enabled \gets False$\;
    }
    \ElseIf{$\textit{is\_exiting}(t)$}{
        $T \gets T\ \textbackslash\ t$\;
    }
    \Else{
        $t.enabled \gets True$\;
        $T \gets \textit{update\_task\_weights}(T,\ d,\ t)$\;
    }
    $\textit{update\_fairness\_bound}(t)$\;
    $T' \gets \textit{get\_enabled\_tasks}(d,\ T)$\;
    $t' \gets \textit{pick\_highest\_weight\_task}(T')$\;
    $next\_t \gets  \textit{dispatch\_and\_run}(t')$\;
    $S \gets \textit{append}(S, t'.next\_event)$\;
    $t \gets next\_task()$\;
}
\end{algorithm}

%% file: figures/algo_mutation.tex
\begin{algorithm}[htbp]
\small
\caption{Concurrency Mutation}
\label{alg:mutation}
\DontPrintSemicolon
\SetKwInput{Input}{Input}
\SetKwInput{Output}{Output}

\Input{syscall sequence $p$}
\Input{number of syscalls to execute in parallel $N$}
\Output{mutated syscall sequence $P$}

$P \gets empty\mbox{-}list$\;
$S \gets pick\_parallel\_syscalls(N, p)$\;
\If{$S \neq \emptyset$}{
    \For{$i \gets 0$ \KwTo $|p| - 1$}{
        $P \gets append(P, p[i])$\;
        \If{$p[i] \in S$}{
            $s \gets duplicate(p[i])$\;
            $s' \gets annotate\_async(s)$\;
            $P \gets append(P, s')$\;
        }
    }
}
\KwRet{$P$}\;

\end{algorithm}

%% file: sections/05-Implement.tex
\section{Implementation}

We implement the scheduler of \tool as a C/BPF program. 
In total, the \tool scheduler framework is a 790 line of code (LoC) as C/BPF programs, with another 96 LoC in C for the command line utility.
Each scheduling algorithm is implemented as individual C/BPF files, which are included by the \tool scheduler at compile time.
In total, the algorithms together add another 345 lines of code in aggregate to the \tool.
The \tool fuzzing loop builds on top of Syzkaller, changing 358 LoC.
Lastly, we implement the instrumentation as a LLVM pass in 409 LoC in C++ .
The helper function that invokes the scheduler is injected by the pass. It consists of an additional 23 LoC in C, including the necessary preemption admissibility check. We show the simplified snippet of the admissibility check in Listing~\ref{fig:might_sleep}, including those extracted from kernel native assertions \rev{in \texttt{might\_sleep()}}.

The LLVM pass instruments load and store instructions that access cross-thread variables, which are identified in LLVM IR by their synchronization scope (e.g., \texttt{syncScope::System}).
For semantic scheduling algorithms such as POS, the instrumentation sends a message to the scheduler before yielding the CPU.
This message encodes the target memory address and whether the operation is destructive (i.e., a store instruction), using scheduling-policy-specific fields in the task struct.

\input{figures/might_sleep}

%% file: figures/might_sleep.tex
\begin{lstlisting}[
style=Cstyle, 
caption={The simplified snippet of the admissibility check for additional scheduling points inject by \tool.},
label={fig:might_sleep}]
if (current->policy == SCHED_EXT 
  && get_current_state() == TASK_RUNNING 
  && !current->non_block_count 
  && !is_idle_task(current) 
  && preempt_count() == 0 
  && !irqs_disabled() 
  && rcu_preempt_depth() == 0 ) {
      // ...	
      yield();
}
\end{lstlisting}

%% file: sections/06-Eval.tex
\section{Evaluation}
\label{sec:eval}

In this section, we evaluate \tool on recent \rev{\texttt{sched\_ext}-capable Linux kernels} in comparison with the state-of-the-art fuzzers. 

We seek to answer the following research questions:

\begin{itemize}[leftmargin=2em]
  \item \textbf{RQ1:} How does \tool perform in branch coverage?
  \item \textbf{RQ2:} How much overhead is introduced by \tool?
  \item \textbf{RQ3:} How does \tool perform in bug detection and bug triggering?
  \item \textbf{RQ4:} Does \tool achieve our goals of maintainability and extensibility?
\end{itemize}


\subsection{Experiment Setup}

\textbf{Kernel under Test.} \rev{All experiments in this section use Linux 6.13-rc4 to keep the kernel and baseline ports fixed across fuzzers. Our separate real-world bug-finding campaigns run on multiple Linux kernels that support \texttt{sched\_ext}.}  The kernels are compiled using the same compilation configurations recommended by syzbot~\cite{syzbot}, with \texttt{KCOV} enabled for collecting coverage feedback and \texttt{KASAN} enabled for detecting memory \startrev safety violations
that arise from concurrency bugs, such as concurrent Use-After Free bugs.
We omit experiments with other test oracles, including KCSAN, because these are generally orthognal --- test-drivers like \tool generate many different executions and test oracles like KCSAN determine whether a given execution manifested a bug. 
Indeed, many concurrency bugs such as the JFS bug (Figure \ref{fig:jfs-partial}) are not data-races and thus cannot be detected by KCSAN, but rather are an issue with a particular ordering of two critical sections. 
\finishrev
\textbf{Baselines.} 
We compare \tool with Snowboard~\cite{Snowboard}, SegFuzz~\cite{SegFuzz}, and Syzkaller~\cite{Syzkaller}. Snowboard and SegFuzz are state-of-the-art concurrency kernel fuzzers, while Syzkaller is the most widely used sequential-oriented kernel fuzzer.

We manually ported these fuzzers to the recent kernel version. 
Unfortunately, we found Snowboard to have two severe practical limitations: (1) it required a prohibitively large amount of resources for pre-processing, including roughly 100TB of disk space and (2) the pre-processing analysis of potential memory conflicts by Snowboard does not support many newer kernel features, several of which are present in the reproducers for bugs in our benchmark. 
As such, we only compare against Snowboard's throughput and not its bug-finding ability.
We evaluate three CCT algorithms in \tool: \toolrw, \toolpos, \toolpct as well as an ablative baseline, \toolminus.
The baseline \toolminus consists of \toolrw without any additional injected scheduling points (c.f. Section \ref{sec:inst}). We exclude the RP algorithm as POS is a refinement of it. Unless otherwise noted, \tool refers to the default algorithm of RW.
In addition, all the fuzzers use the same seed corpus according to \cite{syz_corpus}.

\textbf{Experimental Parameters.} For each fuzzing experiment, we conduct 10 trials of 48 hours. 
For a fair comparison, we configure Syzkaller, SegFuzz and \tool with the same resources, including 2 cores and 2 GB of memory for each VM. We ran 4 VMs each with 2 fuzzing processes for all experiments with SECT, Syzkaller and Snowboard. For SegFuzz, we ran with the same compute resources (CPU/RAM) but twice as many VM instances (to offset the limitation of 1 SegFuzz process per VM).

\textbf{Hardware.} All experiments were conducted on a 64-bit Ubuntu 22.04 LTS machine with Intel Xeon Platinum 8468V (96 physical cores) and 504 GB of RAM.


\subsection{Code Coverage and Performance Overhead}
\label{sec:eval-cov}


To answer \textbf{RQ1}, we monitor the fuzzing process and record the branch coverage achieved by Syzkaller, SegFuzz and \tool.
Figure~\ref{fig:cover} illustrates the comparison of branch coverage obtained by each fuzzer.   \tool achieves 38\% higher branch coverage than SegFuzz. 
In comparison to the sequential-oriented fuzzer Syzkaller, \tool and SegFuzz reach 88\% and 63\% of its branch coverage, respectively.

To analyze this reduction in coverage and respond to \textbf{RQ2}, we also examine the throughput and latency of each fuzzer, as detailed in
 \autoref{tab:throughput}.
In terms of the latency per test-case, \tool 's execution overhead for a given syscall program is only 11\% more than that of the sequential fuzzer Syzkaller. Critically, this represents a 72\% reduction in latency compared to existing concurrency fuzzers on average.
We consulted the authors of Snowboard and they confirmed that a significant drop in performance was expected for their fuzzer as a result of the heavyweight control it imposes at the hypervisor level. In contrast, \tool’s kernel-native approach avoids these expensive mechanisms and cross-layer interventions.
While \tool’s total throughput is approximately half that of Syzkaller, it vastly outperforms other concurrency fuzzers. SegFuzz achieves only 9\% of \tool's throughput. A primary bottleneck for SegFuzz is how it enforces scheduling control: it is restricted to a single test instance per VM due to a reliance on a limited number of hardware breakpoints.
In contrast, Syzkaller and \tool are able to run multiple parallel instances per VM, providing better scalability (\textbf{C1}).

\input{figures/cover}
\input{figures/throughput-latency}

We explain \tool's coverage reduction compared with Syzkaller with three factors. First, concurrency fuzzers allocate a portion of the computing resources to exploring thread interleavings, which generally contributes little to branch coverage. As a result, fewer resources are available for sequential fuzzing, leading to reduced coverage. Second, the CCT scheduling for tasks and the kernel instrumentation for injecting scheduling points incur additional overhead for kernel execution and thus reduces the throughput. Third, \tool has a higher chance to trigger concurrency issues (discussed in Section~\ref{sec:bug-bench}), which causes a costly VM reboot and also diminishes the throughput.

\subsection{Scheduling Effectiveness}
\label{sec:bug-bench}
To answer \textbf{RQ3}, we construct a ground-truth benchmark consisting of known concurrency bugs in the Linux kernel.
We collect these bugs by referencing prior work \cite{han2024cardshark} or analyzing Linux commit messages in the past two years based on heuristic rules (e.g., \texttt{\^{ }(?=.syzbot)(?=.race).*}). We excluded entries that lacked an associated patch or PoC, manually ported the remaining bugs to our evaluation kernel, and verified their reproducibility.
On the groundtruth, we evaluate two key capabilities: (1) the number of concurrency bugs discovered within a given time budget, which reflects the overall bug-finding capability of the fuzzers, and (2) the number of attempts required to reliably trigger a concurrency bug from a given sequential input, which highlights the effectiveness of the temporal isolation based CCT scheduler.
Accordingly, we evaluate \tool's scheduling effectiveness in two settings: when integrated with fuzzing campaign to search the input and interleaving space (bug-finding) and when used for scheduling independently (bug-reproduction).

\input{figures/known-fuzzing}

\subsubsection{Results: Bug-Finding}
\label{sec:bug-finding}

In this first setting, we evaluate the efficacy of \tool in bug detection by conducting a systematic fuzzing campaign on the established benchmark.
At the end of the time budget, we identify the unique bugs discovered and record their cumulative occurrences across all experimental trials.
We present the results of this experiment in \autoref{tab:bench-fuzz}.
During a 48-hour period,  \tool successfully identify up to eight of the ten benchmark bugs, with the performance varying slightly depending on the specific exploration strategy employed.
In contrast, the baseline fuzzers exhibit significantly lower detection capabilities. Syzkaller, SegFuzz and \toolminus find only five, one and six bugs, respectively. 
Two bugs in the benchmark remain undiscovered by all fuzzers.

Based on these results, \tool demonstrates a superior capacity for uncovering concurrency bugs compared to Syzkaller and SegFuzz. While certain cases exhibit a higher variance in discovery frequency (e.g., bug \#3, $p \leq 0.018$)\footnote{Using the chi-squared test~\cite{Pearson1900} with $\alpha = 0.05$ for \tool\_PCT and Syzkaller}, this is a side-effect of the intrinsic throughput trade-off in concurrency fuzzing.
\tool allocates a significant portion of its execution budget to the exploration of the interleaving space. 
While this focus allows it to find  more hard-to-reach concurrency bugs quickly, it does reduce the total volume of generated \emph{inputs} which can reduce consistency when searching both the input and interleaving space.
Indeed, as we will discuss in Section~\ref{sec:bug-repro}, the \tool scheduler is significantly more effective than the native  scheduler at exposing buggy interleavings.
Thus the variation in the consistency stems from differences in \emph{throughput}. 
In other words, \tool pays a cost in terms of consistency of input-space fuzzing for increasing focus on the interleaving-space, though it is still overall a net benefit for concurrency bug-finding on our benchmark.
Finally, SegFuzz significantly underperforms in the benchmark, finding only one bug in all trials.
We directly consulted with the authors of SegFuzz to ensure an optimal configuration.
We attribute this result to SegFuzz’s inherently low execution throughput (Table \ref{tab:throughput}).

\textbf{Scheduling Point Ablation Study.}
To assess how the additional scheduling points impact \tool's effectiveness, we conduct an ablation study with \toolminus, a variant of \tool that relies solely on the kernel’s natural scheduling points without the additional preemption instrumentation.
Without the additionally injected scheduling points, \toolminus discovers fewer bugs overall than full \tool.
However, even with the reduced scheduling granularity, \toolminus is still able to uncover one bug (\#2) that Syzkaller fails to detect in all trials.
At the same time, the limited scheduling points significantly constrain its ability to expose bugs that require precise interleavings (e.g., bug \#2 and \#8). 
These results demonstrate the effectiveness of \scheduler and highlight the benefits of injecting additional scheduling points to enable finer-grained exploration.

\input{figures/survival}

\subsubsection{Results: Bug-Reproduction}
\label{sec:bug-repro}
In addition to the bug-finding setting, we further evaluate the effectiveness of \tool scheduler in exploring the thread interleaving space.
Specifically, we feed the scheduler of \tool with the PoC syscall programs belonging to the benchmark, which are able to trigger the corresponding concurrency bugs.
For each bug, we record the number of execution attempts required to induce a kernel crash.
The native OS scheduler, utilized by existing fuzzers such as Syzkaller, serves as the baseline for this comparison.
Bugs without an available syscall program are omitted in this experiment.
We utilize Kaplan-Meier survival curves~\cite{kaplan1958nonparametric} to visualize the reproduction efficiency for each scheduler in \autoref{fig:scheds-to-bug}.
These plots show the probability that a bug has \emph{not been found yet} (y-axis) after a certain number of schedules have been explored (x-axis).
Curves that exhibit a steeper decline and reach zero more rapidly indicate superior bug-triggering capabilities.

The results demonstrate that, while the number of schedules required to expose concurrency bugs varies significantly, the temporal isolation scheduler of \tool across all implemented algorithms can reproduce these bugs in consistently fewer attempts than the native OS scheduler.
On average, \tool achieves an 11.4$\times$ speed-up in bug reproduction.
This advantage is particularly evident in cases such as bugs \#1 and \#7. For instance, while the native OS scheduler requires up to 13,000 schedules to trigger the bug \#1, \tool consistently exposes the vulnerability in fewer than 500 schedules.
In addition, we find that the PCT algorithm has a worse performance than the RW algorithm in some cases, such as bug \#1 and bug \#7 ($p < 0.005$)\footnote{Using the log-rank test~\cite{mantel1966evaluation} with $\alpha = 0.05$}, whereas the RW and POS algorithms exhibit comparable effectiveness on the benchmark.
This discrepancy may be due to the tendency of the PCT algorithm to induce severe starvation of some tasks.
While these algorithms are often effective for user-space program testing, kernel-level tasks may rely on stronger implicit fairness assumptions. Violating these assumptions may lead to redundant or unproductive error paths. These findings suggest that incorporating more sophisticated, fairness-aware scheduling algorithms \cite{zhao2025surw} represents a promising direction for future work.

\subsection{Qualitative Examination of \tool}

\begin{table}[htbp]
\centering
\small
\caption{LoC for four representative CCT scheduling algorithms implemented in \tool.}
\label{tab:algo-impl}
\setlength{\tabcolsep}{4mm}{
\begin{tabular}{ll}
\toprule
\textbf{Algorithm} & \textbf{LoC.} \\ \midrule
PCT~\cite{PCT} & 216 \\
POS~\cite{yuan2018partial} & 66 \\
RW & 40 \\
RP & 23 \\
\bottomrule
\end{tabular}
}
\end{table}

To answer \textbf{RQ4}, we examine several aspects of \tool's implementation.
\rev{While maintenance burden \textbf{(C2)} is difficult to quantify directly, a major source of incompatibility in prior work stems from modifications to the kernel source code.}
Such changes can result in merge conflicts when ported to newer kernel versions or lead to unexpected functionality failures as kernel evolves.
For example, Krace~\cite{Krace} requires over 10k LoC of kernel code modifications.
SegFuzz~\cite{SegFuzz} not only introduces over 5k LoC changes to the VM guest kernel, but also requires additional invasive patches for the \emph{host kernel} as well! In contrast, \tool leverages the compiler infrastructure for kernel instrumentation and adds only a single helper function to the kernel source, comprising fewer than 25 LoC. This is even fewer than Razzer (40 LoC changed) \cite{Razzer} which provides only limited scheduling control.
As a result, \tool substantially reduces the likelihood of merge conflicts when adapting to newer kernel versions.
\rev{This does not guarantee that \tool will remain compatible with all future releases of the Linux kernel without maintenance. However, \texttt{sched\_ext} has remained stable for more than two years, and \tool avoids the invasive kernel-source patches that make prior tools difficult to port across kernel versions. Even if some APIs change, the migration cost is therefore expected to be substantially lower than for patch-heavy approaches.}

For evaluating extensibility \textbf{(C3)}, LoC can serve as a meaningful proxy for the implementation complexity of new scheduling strategies. 
To demonstrate the modularity of \tool's strategies, Table \ref{tab:algo-impl} details the LoC required to implement each of the four representative CCT scheduling algorithms within the framework.
Most of these scheduling algorithms can be implemented in less than 100 LoC, reflecting a relatively low barrier for extending with diverse exploration strategies. This indicates that the goal of extensibility (\textbf{C3}) has been achieved.
The implementation of the PCT algorithm involves additional complexity due to the logic required to mitigate busy-wait loops, which are common in kernel code but lead to pathological behavior for highly unfair scheduling algorithms such as PCT.

%% file: figures/cover.tex
\begin{figure}[t!]
\centerline{\includegraphics[width=0.85\columnwidth]{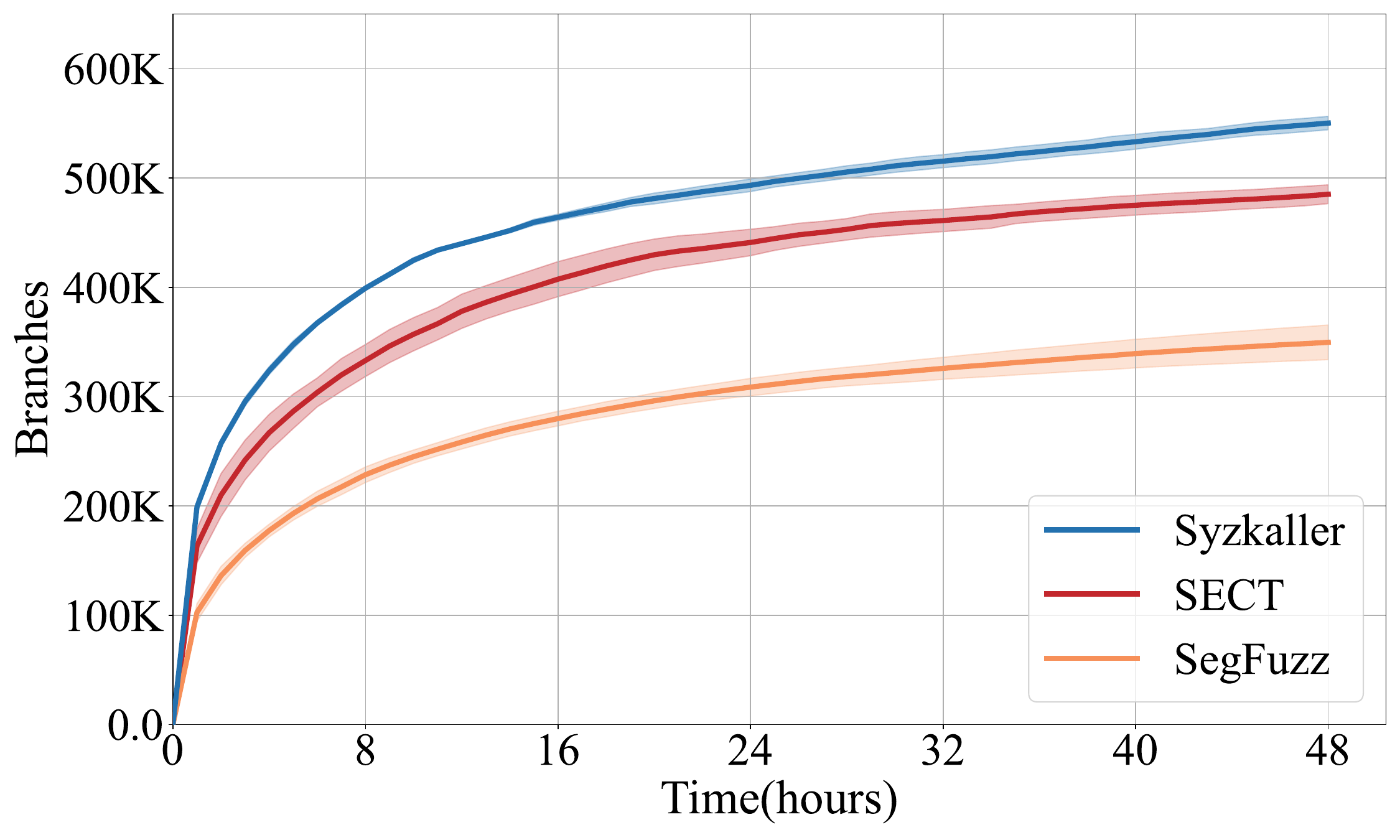}}
\caption{Code coverage achieved by Syzkaller, SegFuzz and \tool (N=10).}
\label{fig:cover}
\end{figure}

%% file: figures/throughput-latency.tex

\begin{table}[t!]
\centering
\small
\caption{Average fuzzing throughput and concurrent testcase-latency over 48 hours (N=10).}
\label{tab:throughput}
\begin{tabular}{@{}cccc@{}}
\toprule
\textbf{Type} &
  \textbf{Fuzzer} &
  \textbf{\begin{tabular}[c]{@{}c@{}}Exec. Time\\ (ms/test)\end{tabular}} &
  \textbf{\begin{tabular}[c]{@{}c@{}}Throughput\\ (\#tests)\end{tabular}} \\ \midrule
\begin{tabular}[c]{@{}c@{}}Sequential\\ Oriented\end{tabular}                  & Syzkaller & 89   & 8.3M  \\ \midrule
\multirow{3}{*}{\begin{tabular}[c]{@{}c@{}}Concurrency\\ Oriented\end{tabular}} & Snowboard & 2498 & -     \\ \cmidrule(l){2-4} 
                                                                                         & SegFuzz   & 228  & 0.38M \\ \cmidrule(l){2-4} 
                                                                                         & \tool      & 99   & 4.4M  \\ \bottomrule
\end{tabular}
\end{table}

%% file: figures/known-fuzzing.tex
\begin{table*}[htbp]
\small
\centering
\caption{Known concurrency bugs found by SegFuzz, Syzkaller, \tool (N=20). Note that 
\toolrw is default \tool.}
\label{tab:bench-fuzz}
\begin{tabular}{@{}ccccccccc@{}}
\toprule
\textbf{ID} & \textbf{Vulnerability} & SegFuzz & Syzkaller & \toolminus & \toolrw & \toolpos & \toolpct \\ \midrule
1  & CVE-2023-31083 \cite{cve-2023-31083} & - & -            & - & -            & -            & -            \\
2  & CVE-2024-42111 \cite{cve-2024-42111} & - & -            & \CIRCLE~~(3) & \CIRCLE~~(8) & \CIRCLE~~(9) & \CIRCLE (10) \\
3  & CVE-2024-44941 \cite{cve-2024-44941} & - & \CIRCLE (10) & \CIRCLE (12) & \CIRCLE~~(1) & \CIRCLE~~(2) & \CIRCLE~~(3) \\
4  & CVE-2024-49903 \cite{cve-2024-49903} & - & \CIRCLE (13) & \CIRCLE (11) & \CIRCLE~~(7) & \CIRCLE (10) & \CIRCLE~~(9) \\
5  & CVE-2024-50125 \cite{CVE-2024-50125} & - & \CIRCLE~~(4) & \CIRCLE~~(5)& \CIRCLE~~(1) & \CIRCLE~~(2) & \CIRCLE~~(1) \\
6  & CVE-2024-57900 \cite{CVE-2024-57900} & - & -            & - & \CIRCLE~~(1) & -            & -            \\
7  & cb2239c1 \cite{cb2239c1}             & - & -            & - & -            & -            & -            \\
8  & 61179292 \cite{61179292}             & - & -            & - & \CIRCLE~~(1) & \CIRCLE~~(4) & \CIRCLE~~(1) \\
9  & 3b9bc84d \cite{3b9bc84d}             & \CIRCLE (15) & \CIRCLE (16) & \CIRCLE (18) & \CIRCLE (16) & \CIRCLE (19) & \CIRCLE (19) \\
10 & 88b1afbf \cite{88b1afbf}             & - & \CIRCLE (15) & \CIRCLE (12) & \CIRCLE (12) & \CIRCLE (18) & \CIRCLE(15)    \\ \midrule
   & Total                                & 1 & 5            & 6 & 8            & 7            & 7            \\ \bottomrule
\end{tabular}
\end{table*}

%% file: figures/survival.tex
\begin{figure*}[htbp]
\centering{
\includegraphics[width=\textwidth]{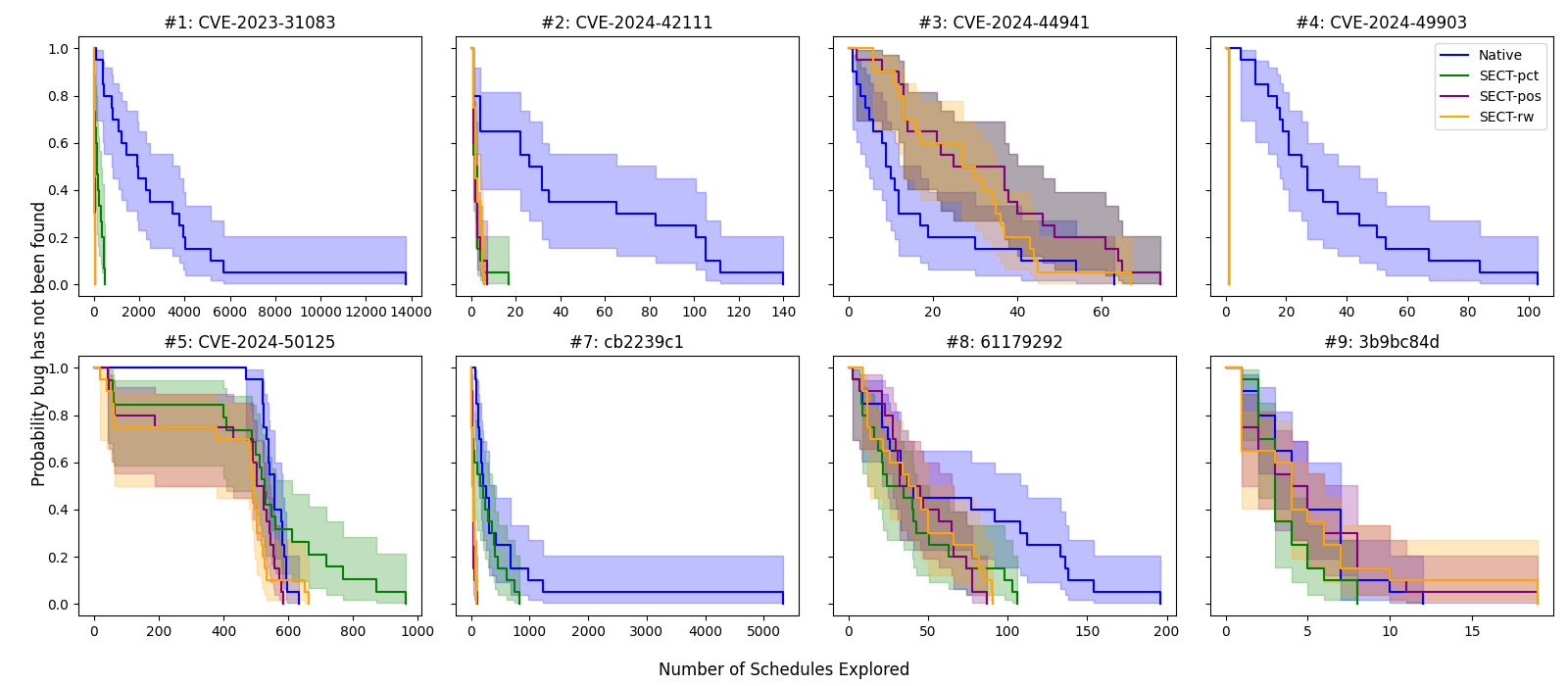}
}
\caption{Cumulative probability of failing to detect the bug relative to the number of explored schedules. Lower values indicate higher efficiency. (N=20).}
\label{fig:scheds-to-bug}
\end{figure*}

%% file: sections/07-Case-study.tex
\section{Real-world Bug Finding}

\input{figures/new-bugs}

\rev{We evaluate \tool across multiple \texttt{sched\_ext}-capable Linux kernels along with the development of \texttt{sched\_ext}.} Specifically, we conducted four independent 7-day fuzzing campaigns. Each campaign was provisioned with four VMs, each allocated 2 CPUs and 2GB of RAM. \rev{These campaigns produced roughly 500 raw crash reports}. For all encountered crashes, we performed a systematic triage procedure consisting of: (1) searching on the internet and \rev{syzbot \cite{syzbot}} to identify and discard duplicate reports; (2) minimizing the collected syscall programs and excluding non-concurrency bugs; (3) performing manual analysis to validate the issue.
\rev{Most reports that did not become new bug reports were duplicates of known issues and the remaining discarded reports were either non-reproducible or unrelated to concurrency.}
Through this process, we have identified eight previously unknown concurrency-related bugs in the Linux kernel, as detailed in \autoref{tab:new-bugs}. Among these bugs, five have already been fixed, and one has been confirmed by kernel developers. The remaining two bugs are currently under review.
We present two representative case studies of these bugs that have been fixed or confirmed by developers, illustrating the practical effectiveness of our approach:

\textbf{Case Study \uppercase\expandafter{\romannumeral 1}: use-after-free in \texttt{sl\_sync}.} This bug is a concurrent use-after-free in the \texttt{sl\_sync} function.
It is caused by a kernel thread improperly freeing a Serial Line IP Device (\texttt{slip\_dev}) without first acquiring an \texttt{rtnl\_lock}.
The use occurs when attempting to clean up existing open SLIP channels.
Average number of trials to reproduce this bug appears in Figure~\ref{fig:scheds-to-bug-slip}.
\tool reproduces this bug in an order of magnitude fewer schedules than the native OS scheduler.

\input{figures/survival-slip-open}


\textbf{Case Study \uppercase\expandafter{\romannumeral 2}: deadlock in \texttt{pernet\_ops\_rwsem}.} Our development of \tool also lead to improvements in the \texttt{sched\_ext} framework. 
\autoref{motiv:deadlock} shows a deadlock bug discovered by \tool in the \texttt{sched\_ext} subsystem of a fork of the Linux kernel~\cite{sched_ext}.
This deadlock could be triggered if a task takes \texttt{cpu\_hotplug\_lock} between \texttt{scx\_pre\_fork()} and \texttt{scx\_post\_fork()}.
One possible unsafe locking scenario is depicted: CPU0 is currently holding the lock \texttt{pernet\_ops\_rwsem}, and requires the \texttt{cpu\_hotplug\_lock}. However, CPU1 already holds \texttt{cpu\_hotplug\_lock}, and requires the \texttt{pernet\_ops\_rwsem} lock that CPU0 holds. Since both CPUs are waiting on each other and have no way to preempt the resource, a deadlock scenario is reached.

\input{figures/deadlock}

%% file: figures/new-bugs.tex
\begin{table*}[h]
\centering
\small
\caption{Previously unknown concurrency-related bugs in Linux kernel discovered by \tool.}
\label{tab:new-bugs}
\begin{tabular}{@{}cccccc@{}}
\toprule
\textbf{ID} & \textbf{Version} & \textbf{Bug Type}   & \textbf{Location} & \textbf{Subsystem} & \textbf{Status} \\ \midrule
1   & 6.13-rc4   & use-after-free read      & sl\_sync          & driver/net/slip   & confirmed       \\
2   & 6.13-rc4   & use-after-free read      & netdev\_walk\_all\_lower\_dev & net/core & \rev{fixed$^{\dagger}$} \\
3   & 6.13-rc4   & null-ptr-deref           & jfs\_ioc\_trim    & fs/jfs            & fixed        \\
4   & 6.13-rc4

& deadlock  &  nr\_del\_node & net/netrom & \rev{confirmed$^{\dagger}$} \\
5   & sched\_ext   & RCU anno missing &  xa\_get\_order & kernel/mm & fixed \\
6   & sched\_ext   & deadlock &  pernet\_ops\_rwsem &  kernel/sched & fixed \\
7   & sched\_ext   & memory leak &  scx\_ops\_enable & kernel/sched & fixed \\
8   & sched\_ext   & deadlock   & lockdep\_assert\_rq\_held & kernel/sched & fixed \\
\bottomrule
\end{tabular}
\end{table*}

%% file: figures/survival-slip-open.tex
\begin{figure}[h]
\centering{
\includegraphics[width=0.88\columnwidth]{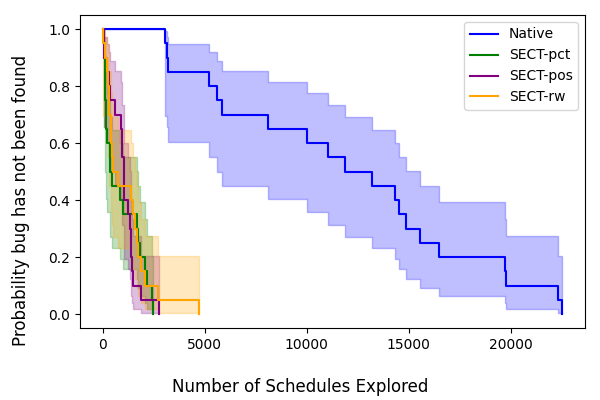}
}
\caption{Kaplan-Meier survival analysis for the  confirmed use-after-free bug in \texttt{sl\_sync} identified by \tool (N=20).}
\label{fig:scheds-to-bug-slip}
\end{figure}

%% file: figures/deadlock.tex
\begin{figure}[t!]
\centerline{\includegraphics[width=\columnwidth]{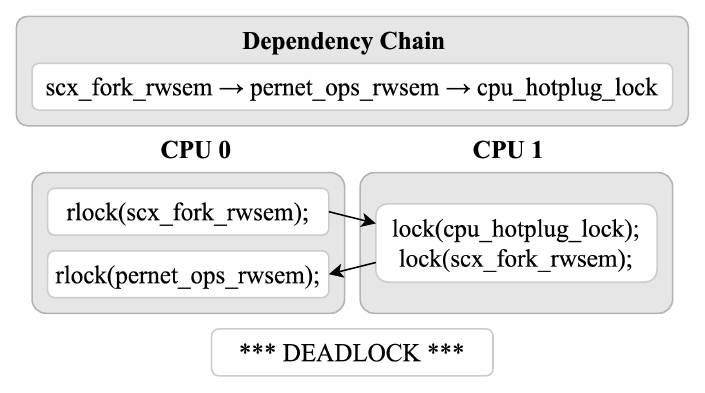}}
\caption{The root cause of a deadlock bug in the \texttt{sched\_ext} class identified by \tool.}
\label{motiv:deadlock}
\end{figure}

%% file: sections/08-Limitations.tex
\section{Limitations}
\label{sec:limit}

\startrev \paragraph{Soundness.} 
Without additional instrumentation, \tool is sound in that it will not emit schedules or report false-positive kernel bugs that are not possible to reproduce without \tool.
However, when using additional instrumentation (c.f. Section \ref{sec:inst}) the runtime admissibility checks extracted from \texttt{might\_sleep()} do not \emph{guarantee} the safety of additionally injected preemptions.
As a result, if the safety checks incorrectly report an unsafe context switch to be safe, it is possible for \tool to report false positive bugs.
However, when evaluating \tool on recent versions of the Linux kernel, we manually triaged more than 500 crash reports and did not see any reports that we could trace to false positives from unsafe preemptions.
Among the non-reported crashes which we manually investigated, 87\% were duplicates of previously reported crashes on \texttt{syzbot}\footnote{\url{https://syzkaller.appspot.com/upstream}}\freefootnote{\rev{\textsuperscript{$\dagger$} Concurrently discovered and reported by syzbot}}, roughly 6\% were not reproducible, and roughly 7\% were not related to concurrency. 
To avoid reporting false-positive bugs, we did verify bugs with the native scheduler when possible (e.g. Figure \ref{fig:scheds-to-bug-slip}), though in many cases this can be prohibitively expensive due to the weak exploratory power of the native scheduler.
However, we note that attempting to reproduce a bug on an uninstrumented kernel is not a definitive false-positive test: failure to reproduce may simply mean the buggy interleaving is too rare to occur naturally, not that the report is spurious.
When we weren’t able to reproduce bugs outside of SECT, we collaborated with kernel developers to ensure bugs were not false positives, though this did not end up being the case for any of our found bugs.

\paragraph{Completeness.}
The CCT approach of schedule serialization and injection of scheduling points is theoretically complete in that it can produce any sequentially consistent schedule which is realizable by the native OS scheduler.
However, \tool makes three practical compromises which may make some behaviors impossible to observe.
For completeness, a scheduling point must be inserted and triggered at runtime between any two visible (non-commuting across threads) operations, so that those operations can be reordered by the scheduler. 
(1) Our LLVM instrumentation pass may miss some of these visible operations in the kernel due to inline assembly code, synchronization primitives we may have failed to identify as instrumentation targets, and core modules or IRQ contexts omitted from instrumentation.
Additionally, (2) the admissibility checks (c.f. Section \ref{sec:inst}) may spuriously mark some safe preemption contexts as unsafe at runtime, resulting in some instrumented scheduling points not being triggered during execution. 
Lastly, (3) \tool uses a compile-time configurable sampling-rate for each additional scheduling point, allowing users to increase performance at the cost of scheduling granularity.  
Sampling may cause \tool to miss bugs in a particular execution, due to not sampling a key scheduling point. 
However, it notably does not affect \tool's \textit{probabilistic completeness}~\cite{zhao2025surw} in that the missed bug would still be observable with some probability in future executions where the key scheduling point \emph{is} sampled.
\finishrev




%% file: sections/09-Perspectives.tex
\section{Conclusion}

\tool is the first kernel-native concurrency fuzzing framework built upon the \textit{scheduler-as-an-explorer} paradigm. It is designed to overcome the architectural bottlenecks of existing kernel concurrency fuzzers by providing fine-grained, lightweight, and extensible scheduling control for CCT. To achieve this, \tool proposes the temporal isolation scheduling with programmable policies, the safe preemption instrumentation mechanism, and the two-phase input-and-concurrency fuzzing workflow. Our evaluation demonstrates that \tool achieves substantial improvements in branch coverage,
efficiency, and bug discovery effectiveness. 
We anticipate that the design of \tool's components will facilitate developers and researchers to leverage it to increase the fundamental robustness of OS kernels.

%% file: sections/Ethics.tex
\section{Ethical Considerations}

\startrev
\paragraph{Stakeholders}
The list of possible stakeholders for this work is large, including not just maintainers and developers of the Linux kernel but also direct and indirect users and vendors of Linux as well. 
For example, historically, high-severity concurrency vulnerabilities such as the Dirty COW bug~\footnote{\url{https://nvd.nist.gov/vuln/detail/CVE-2016-5195}} have affected commercial vendors of Linux distributions, such as Google and RedHat, consumers of those and other distributions, such as Android phone users, enterprises which deploy their applications on Linux servers, such as Amazon.
These stakeholders risk damage to their reputations, expensive availability outages, and/or costly remediations in the case of vendors.
For consumers and enterprise users, concurrency vulnerabilities can also lead to personal or commercial data theft and denial-of-service or even active exploitation of affected devices.
Given that the Linux kernel is part of the most widely deployed operating system in the world (Android) and is the leading operating system family used by enterprises for server deployments, vulnerabilities and bugs in the kernel can affect innumerable number of users, businesses and customers.

\paragraph{Potential Negative Outcomes and Mitigations} 
During the research process, those most directly affected are the kernel maintainers to whom we reported bugs. Our reports take up their time and effort to process, which can be a burden if the reports are of low quality or spurious – which we attempted to minimize by not submitting duplicate reports. Post-publication, we hope SECT can positively affect downstream users by improving kernel security and reliability and also positively affect maintainers and researchers by providing a more robust substrate with which to test the kernel for concurrency bugs (and reproduce them).

It is possible that in reporting these bugs, a malicious actor could use our reports for exploitation of consume or vendor software before the bugs are patched. However, there are several mitigating circumstances: (1) the \scx fork was not widely deployed at the time of reporting (2) the Linux kernel maintainers are trusted – if a maintainer has already been compromised, the security of that module or subsystem is completely compromised as well and (3) we attempted to provide patches and assisted maintainers in debugging and fixing these bugs as quickly as possible, limiting unpatched exposure time. In the case of the JFS bug, our patch was accepted and merged into the Linux kernel. As far as we know, none of the bugs we have reported were actively exploited, but this nonetheless remains a risk of our research. 
There is an additional risk in that we did not escalate all issues to the security team by default. However, we opted to defer to maintainers in this regard as they are experts in their respective modules.

We have made SECT openly available to assist developers and maintainers to find and debug concurrency issues more reliably. However, this comes with the risk that SECT could be used by malicious actors to find and exploit concurrency bugs more easily. We believe the benefits of open-sourcing SECT outweigh this risk.
Lastly, there is a risk of users of SECT inadvertently exposing themselves to exploitation via concurrency bugs. While we believe this risk to be low due to the difficulty of actively exploiting concurrency bugs, we address this by providing instructions for setting up SECT only within a virtualized environment (despite the scheduler being possible to run natively) and documenting risks involved with using it.
\paragraph{Bug Disclosure Process}
We disclosed bugs directly to kernel maintainers and developers to facilitate swift remediation. In the case of the four \scx bugs, we reported them in the developer Slack channel. In the case of the mainline kernel bugs, we reported two of them to relevant trusted maintainers via emails identified by the \texttt{get\_maintainers.pl} script. We inquired directly to maintainers if bug reports required additional escalation to the kernel security team, but in all cases they did not indicate an escalation was necessary. The remaining two bugs were found and reported by syzbot before we had a chance to submit a bug report.

\paragraph{Bug Impacts}
All of the \scx bugs have been fixed by developers after our initial reports. We believe the extent of the impact to have been reliability disruptions for the users and developers of this kernel in all cases.

Our patch for the JFS bug was accepted and merged into the mainline Linux kernel.
We attempted to submit patches for the \texttt{slip\_open} bug, but we weren’t able to find a correct patch due to our lack of familiarity with the subsystem.
The \texttt{netdev\_walk} bug was patched by a maintainer soon after it was reported.
Lastly, the deadlock in 
\texttt{nr\_del\_node} had a valid patch submitted by a maintainer, but this patch has been postponed in favor of a larger refactoring to clean up an error-prone locking pattern.

It is difficult to gauge the impact of the Linux kernel bugs beyond non-deterministic reliability issues for users. However, because of the broad scope of Linux's user-base, even sporadic issues can affect a large number of stakeholders.

\paragraph{Decision to Publish}
We began this project with this goal in mind: to make concurrency debugging and testing of kernel code more tractable. We knew that many of the concurrency bugs currently submitted are done so without reliable reproducers at all. Indeed we believe this to be a contributing factor to why the JFS bug was only partially fixed.

We decided to submit for publication after we were confident in its effectiveness from our experiments, as this meant it could be deployed by others to actively enhance the security and reliability of kernels.
\finishrev

%% file: sections/OpenScience.tex
\section*{Open Science}
To facilitate future research on OS kernel fuzzing, we 
open-source \tool 
at \url{https://github.com/m0ck1ng/sect} and publish a persistent artifact at \url{https://doi.org/10.6084/m9.figshare.32468280}. 
We hope that our contributions
will foster further advancements in kernel security research.

\section*{Acknowledgments}
This research is supported by the National Research Foundation, Singapore, and Cyber Security Agency of Singapore under its National Cybersecurity R\&D Programme (Fuzz Testing $<$NRF-NCR25-Fuzz-0001$>$).
Any opinions, findings and conclusions, or recommendations expressed in this material are those of the author(s) and do not reflect the views of National Research Foundation, Singapore, and Cyber Security Agency of Singapore.

%% file: sections/usenix26-ae.tex

\appendix
\section{Artifact Appendix}

\subsection{Abstract}
This artifact contains \tool, as well as infrastructure for setting it up on a recent target kernel (v6.13-rc4), our benchmark of kernel concurrency bugs used in our evaluation, and data from our real-world bug-finding runs.
\subsection{Description \& Requirements}

Generally, the system requirements for this artifact are (1) a relatively recent version of Ubuntu (tested on 24.04) (2) Docker (3) QEMU with KVM enabled.
In terms of hardware recommend a machine with at least 8 logical CPU cores and 16GB of RAM.
All other setup should be handled by scripts in the \tool repository.

\subsubsection{Security, privacy, and ethical concerns}
The risks of using \tool are minimal, especially if used within a VM, as is the documented setup in our artifact.
If you attempt to run \tool on your host system outside of a VM, any resulting kernel panics will of course affect your host system directly. 

\subsubsection{How to access}
Our artifact is available open-source on Github at
\url{https://github.com/m0ck1ng/sect} and also permanently archived at \url{https://doi.org/10.6084/m9.figshare.32468280}

\subsubsection{Hardware dependencies}
\tool does not require any specific hardware, though we recommend a multi-core system with at least 8 logical cores for performance when running fuzzing campaigns and to ensure that the Linux kernel itself can be rebuilt by your system in a reasonable amount of time.
To reproduce the full fuzzing experiments, we recommend a machine with at least 80 cores and 2GB of RAM per core, to allow all 10 trials to be run in parallel.

\subsubsection{Software dependencies}
We have developed and tested \tool on Ubuntu 24.04.
Additional recent versions of Ubuntu or Debian are also likely to work without modification.
Other Linux distributions may also work, but are untested and will require some slight modifications of shell scripts (e.g. to install dependencies from \texttt{rpm} instead of \texttt{apt}).

Additionally, the host system needs Docker installed to use our provided setup.
Docker is used purely for a hermetic build environment to ensure all other dependencies are the correct versions.
For example, \tool and the kernel under test are mounted into a docker container as a volume, and built inside the container rather than on the host system.

The fuzzing setup uses a forked version of syzkaller, which runs in one or more virtual machines using QEMU.
QEMU is installed as part of the build script automatically on the host system via \texttt{apt}.

Finally, to create a disk image for the guest VM, \texttt{debootstrap} is also a required dependency, installed automatically by the main build script.

\subsubsection{Benchmarks}
We have provided the details of the ten concurrency bugs in the benchmark from our experiments in our artifact, including reproducing inputs and a kernel patch for v6.13-rc4 which introduces all of the bugs into that kernel version. 

\subsection{Set-up}

All that is needed for setup is to clone the \tool repository:\\

\texttt{git clone https://github.com/m0ck1ng/sect.git}\\

And to install Docker:\\

\url{https://docs.docker.com/engine/install/}

\subsubsection{Installation}

We have consolidated the installation into a single script: \\

\texttt{./build.sh}\\

Note that some operations may require superuser permissions to e.g. install dependencies or mount a volume for disk image creation.

\subsubsection{Basic Test}
The best way to check that \tool is working after installation is to start a fuzzing campaign via:\\

\texttt{./syzkaller/bin/syz-manager \\ 
-config configs/syzkaller.cfg.example}\\

The configuration should have been pre-populated by the installation process, but it can be edited according to your system requirements as necessary.
You should see output at the command line and in the web interface (\texttt{http://0.0.0.0:56741}) containing:\\

\texttt{Using thread scheduler: <path>} \\ 

and later you should see that \texttt{schedCollided} increases over time, e.g.:\\

\texttt{VMs 4, executed 2434, schedCollided 650, cover 29526, signal 38453/47876, crashes 0, repro 0, triageQLen 8817}

\subsection{Evaluation workflow}

\subsubsection{Major Claims}

\begin{itemize}
    \item[(C1):] 
    \tool achieves higher fuzzing throughput and coverage that other tools, shown in Figure 3 and Table 2.

    \item[(C2):] \tool is more effective at finding concurrency bugs than the native scheduler under Syzkaller or SegFuzz, shown by Table 3 and Figure 4.

\end{itemize}

\subsubsection{Experiments}

\paragraph{(E1)}
This experiment reproduces the code coverage and throughput results for \tool (C1) as well as the Bug-Finding Results (C2) shown in Table 3.
Please note that for all experiments to finish in 48 hours, all trials must be run simultaneously, requiring at least 80 CPU cores for \tool (4 VM instances * 2 cores per instance * 10 trials).

[6 Hours Human] [48 Hours Compute]
\begin{enumerate}
    \item \textit{Preparation.}Build the benchmark kernel with all 10 concurrency bugs by applying the patch provided in our artifact and building \tool with the provided script.
    \item \textit{Experiment.} Run a fuzzing campaign via \texttt{syz-manager} as outlined in the artifact using the provided configuration for each \emph{trial}, using a timeout of 48 hours.
    \item \textit{Results.} For each crash found by \tool, manually check if it corresponds to any of the bugs in the benchmark. Note that once a bug has been correlated, any crashes from other trials with the same hash can also be mapped to that bug.
    We compute throughput and latency from the total number of executions reported by \texttt{syz-manager} and the 48-hour time period.
    To obtain coverage data we use the documented utility provided by Syzkaller: \url{https://github.com/google/syzkaller/blob/master/docs/coverage.md}
\end{enumerate}

We note that for comparing with Syzkaller, most of the above steps are identical, save for using a clean build of Syzkaller in lieu of \tool.
For SegFuzz, we use the instructions provided by the tool authors. 
For Snowboard, we similarly use the instructions provided by the tool authors.

\paragraph{(E2)}
This experiment reproduces the survival plots in Figure 4 (C2). 
Unfortunately, we didn't find a good way to automate these experiments, so they are currently interactive and thus take a very long time to reproduce.

[>10 Hours Human + Compute (interactive)]
\begin{enumerate}
    \item \textit{Preparation.} Build the benchmark kernel with all 10 concurrency bugs by applying the patch provided in our artifact and building \tool with the provided script.
    \item \textit{Experiment.} 
    Start a VM instance using the script provided in our artifact \texttt{./run\_qemu.sh}.
    Copy all necessary binaries and programs into the guest with the provided script \texttt{./copy.sh}.
    Use SSH to obtain a second terminal window inside the guest.
    In one window, start the sched-ext scheduler \texttt{./scx\_serialise} with the desired scheduling algorithm via \texttt{-r}.
    In the other window, execute the desired reproducer with \texttt{syz-execprog} using two processes and 10000 repetitions.
    \item \textit{Results.} For each crash, we record the number of schedules until the crash as logged by the window with \texttt{scx\_serialise}. 
    We generate the survival plots using the script provided in our artifact (\texttt{scripts/analysis/plot.py}).
\end{enumerate}


\subsection{Version}
Based on the LaTeX template for Artifact Evaluation V20231005. Submission,
reviewing and badging methodology followed for the evaluation of this artifact
can be found at \url{https://secartifacts.github.io/usenixsec2026/}.